\documentclass[prd,twocolumn,showpacs,showkeys,preprintnumbers,floatfix,
  nofootinbib,superscriptaddress]{revtex4-1}
\usepackage{amsfonts} % AMS
\usepackage{amssymb} % AMS
\usepackage{amsmath} % AMS
\usepackage{subfigure} 
\usepackage{graphicx} % Include figure files
\usepackage{array} % array
\usepackage{dcolumn} % Align table columns on decimal point
\usepackage{bm} % bold math
\usepackage{latexsym} % latex symbols
\usepackage{longtable} % long tables
\usepackage{hyperref} % hypertext links 
\usepackage{verbatim}
\graphicspath{{./figs/}}
\begin{document}
\title{Improved determination of $B_K$ with staggered quarks}
\author{Taegil Bae}
\affiliation{
  Lattice Gauge Theory Research Center, FPRD, and CTP, \\
  Department of Physics and Astronomy,
  Seoul National University, Seoul, 151-747, South Korea
}
\author{Yong-Chull Jang}
\affiliation{
  Lattice Gauge Theory Research Center, FPRD, and CTP, \\
  Department of Physics and Astronomy,
  Seoul National University, Seoul, 151-747, South Korea
}
\author{Hwancheol Jeong}
\affiliation{
  Lattice Gauge Theory Research Center, FPRD, and CTP, \\
  Department of Physics and Astronomy,
  Seoul National University, Seoul, 151-747, South Korea
}
\author{Chulwoo Jung}
%
%\email[E-mail: ]{chulwoo@bnl.gov}
%
\affiliation{
  Physics Department, Brookhaven National Laboratory,
  Upton, NY11973, USA
}
\author{Hyung-Jin Kim}
\affiliation{
  Physics Department, Brookhaven National Laboratory,
  Upton, NY11973, USA
}
\author{Jangho Kim}
\affiliation{
  Lattice Gauge Theory Research Center, FPRD, and CTP, \\
  Department of Physics and Astronomy,
  Seoul National University, Seoul, 151-747, South Korea
}
\author{Jongjeong Kim}
\affiliation{
  Lattice Gauge Theory Research Center, FPRD, and CTP, \\
  Department of Physics and Astronomy,
  Seoul National University, Seoul, 151-747, South Korea
}
\author{Kwangwoo Kim}
\affiliation{
  Lattice Gauge Theory Research Center, FPRD, and CTP, \\
  Department of Physics and Astronomy,
  Seoul National University, Seoul, 151-747, South Korea
}
\author{Sunghee Kim}
\affiliation{
  Lattice Gauge Theory Research Center, FPRD, and CTP, \\
  Department of Physics and Astronomy,
  Seoul National University, Seoul, 151-747, South Korea
}
\author{Weonjong Lee}
%
%\email[E-mail: ]{wlee@snu.ac.kr}
%
%\altaffiliation[Visiting professor at ]{
%  Physics Department,
%  University of Washington,
%  Seattle, WA 98195-1560, USA
% }
%
\affiliation{
  Lattice Gauge Theory Research Center, FPRD, and CTP, \\
  Department of Physics and Astronomy,
  Seoul National University, Seoul, 151-747, South Korea
}
\author{Jaehoon Leem}
\affiliation{
  Lattice Gauge Theory Research Center, FPRD, and CTP, \\
  Department of Physics and Astronomy,
  Seoul National University, Seoul, 151-747, South Korea
}
\author{Jeonghwan Pak}
\affiliation{
  Lattice Gauge Theory Research Center, FPRD, and CTP, \\
  Department of Physics and Astronomy,
  Seoul National University, Seoul, 151-747, South Korea
}
\author{Sungwoo Park}
\affiliation{
  Lattice Gauge Theory Research Center, FPRD, and CTP, \\
  Department of Physics and Astronomy,
  Seoul National University, Seoul, 151-747, South Korea
}
\author{Stephen R. Sharpe}
%
%\email[E-mail: ]{sharpe@phys.washington.edu}
%
\affiliation{
  Physics Department,
  University of Washington,
  Seattle, WA 98195-1560, USA
}
\author{Boram Yoon}
\affiliation{
  Los Alamos National Laboratory,
  Theoretical Division T-2, MS B283,
  Los Alamos, NM 87545, USA 
}
\collaboration{SWME Collaboration}
\date{\today}
\begin{abstract}
We present results for the kaon mixing parameter $B_K$ 
obtained using improved staggered fermions on 
a much enlarged set of MILC asqtad lattices.
Compared to our previous publication, which was based largely
on a single ensemble at each of the three lattice spacings 
$a\approx 0.09\;$fm, $0.06\;$fm and $0.045\;$fm,
we have added seven new fine
and four new superfine ensembles, with a range of values of
the light and strange sea-quark masses.
We have
also increased the number of measurements on one of the original ensembles.
This allows us to do controlled extrapolations in the
light and strange sea-quark masses, which we do simultaneously
with the continuum extrapolation.
This reduces the extrapolation error and improves the reliability
of our error estimates.
Our final result is 
$\hat{B}_K = 0.7379 \pm 0.0047 (\text{stat}) \pm 0.0365 (\text{sys})$.
\end{abstract}
\maketitle

\section{Introduction}
\label{sec:intr}

The kaon B-parameter, $B_K$, is one of the important hadronic inputs
into the unitary triangle analysis of flavor physics.  Fully
controlled, first-principles calculations are only available using
lattice QCD, and consistent results utilizing several fermion
discretizations have now been obtained \cite{ Arthur:2012opa,
  Bae:2011ff, Durr:2011ap, Aubin:2009jh}.
%
% EDIT
%
For a recent review, see Ref.~\cite{Aoki:2013ldr}.  Among these
results are those we have previously presented using improved
staggered fermions for both valence and sea
quarks~\cite{Bae:2010ki,Bae:2011ff} Here we provide a significant
update of these results, in which the control of several sources of
systematic error is markedly improved.

Our previous result (Ref.~\cite{Bae:2011ff}) was based
largely on a single gauge ensemble at each of three lattice spacings
($a\approx 0.045$, $0.06$ and $0.09\;$fm---dubbed ``ultrafine'', ``superfine''
and ``fine'' hereafter).
These ensembles had approximately the same physical values for
the light sea-quark mass ($m_\ell$, the degenerate up and down mass)
and the strange sea-quark mass ($m_s$), allowing a continuum extrapolation.
Extrapolating $m_\ell$ and $m_s$ to their physical values was not, however,
possible using these ensembles.
Instead, we used 
results from ``coarse'' ensembles ($a\approx 0.12\;$fm),
for which a range of values of $m_\ell$ was available,
to argue that the extrapolations in $m_\ell$ and $m_s$ 
would lead to only a small shift in $B_K$.
These ensembles were too coarse, however, to be included in our
continuum extrapolation.

The main improvements since Ref.~\cite{Bae:2011ff} are the inclusion of
many additional ensembles and an increase in the number of measurements on 
ensembles used previously.
Specifically, we have added results on seven more fine 
and four more superfine ensembles, while more than quadrupling the
number of measurements on the previously used superfine ensemble.
(See Table~\ref{tab:milc:para} below.)
The main impact of these improvements is that we can now do a controlled
extrapolation in $m_\ell$ and $m_s$, which we do simultaneously
with the continuum extrapolation.
This leads to better understood and, in most cases,
numerically smaller systematic errors.
It also removes the need to use the coarse ensembles,
with the entire analysis now carried out using results from
the three finest lattice spacings.

In this report, we refer to Refs.~\cite{Bae:2010ki} and \cite{Bae:2011ff}
for explanations of many technical details.
Since those papers were published, several updates have
appeared in conference proceedings,
the most recent being Ref.~\cite{Bae:2013lja}. This report is based on
our final data set, and the result supersedes earlier ones.

\section{Data sample}
\label{sec:meas}
The kaon B parameter $B_K$ is defined as 
\begin{equation}
B_K(\mu,\text{R}) = 
\dfrac{\langle K^0 | O_{\Delta S = 2}(\mu,\text{R}) | \bar{K}^0 \rangle}
{8f_K^2 M_K^2/3}
\label{eq:bkdef}
\end{equation}
where $R$ is the renormalization scheme in which the operator 
$O_{\Delta S = 2} = \sum_{\nu} [ \bar{s} \gamma_\nu (1 -
  \gamma_5) d ] [ \bar{s} \gamma_\nu (1 - \gamma_5) d ]$
is defined,
with $\mu$ is the corresponding renormalization scale.
The standard scheme used in phenomenology is the $\overline{\rm MS}$
scheme with naive dimensional regularization (NDR) for $\gamma_5$.  We
match our lattice-regulated operators to this scheme, usually called
NDR, using one-loop matching factors from Ref.~\cite{KLS2}.
We use the ensembles generated by the MILC collaboration with
$N_f=2+1$ flavors of asqtad staggered sea quarks and a Symmanzik-improved
gauge action~\cite{Bazavov:2009bb}.
For valence quarks, we use HYP-smeared staggered 
fermions~\cite{Hasenfratz:2001hp}.
The advantages of this mixed action set-up are explained in 
Ref.~\cite{Bae:2010ki}.
%

%-----------------------
% MILC asqtad ensembles
%-----------------------
%
%
%
\begin{table}[tb!]
\caption{MILC asqtad ensembles used to calculate $B_K$.
$a m_\ell$ and $a m_s$ are the masses, in lattice units,
of the light and strange sea quarks, respectively. ``ens'' indicates
the number of configurations on which ``meas'' measurements are made.
Ensembles added since Ref.~\cite{Bae:2011ff} are denoted \texttt{new}
while that with improved statistics is denoted \texttt{update}.
Note that the numbering of
the ID tags for fine and superfine lattices 
does not follow the ordering of $a m_\ell$.
}
\label{tab:milc-lat}
\begin{ruledtabular}
\begin{tabular}{ c | c | c | c | c | c }
%%%\hline\hline
$a$ (fm) & $am_l/am_s$ & geometry & ID & ens $\times$ meas
%%%& $B_K$($\mu=2$ GeV) 
& status \\
\hline
0.12 & 0.03/0.05  & $20^3 \times 64$ & C1 & $564 \times 9$ & old \\
0.12 & 0.02/0.05  & $20^3 \times 64$ & C2 & $486 \times 9$ & old \\
0.12 & 0.01/0.05  & $20^3 \times 64$ & C3 & $671 \times 9$ & old \\
0.12 & 0.01/0.05  & $28^3 \times 64$ & C3-2 & $275 \times 8$ & old \\
0.12 & 0.007/0.05 & $20^3 \times 64$ & C4 & $651 \times 10$ & old \\
0.12 & 0.005/0.05 & $24^3 \times 64$ & C5 & $509 \times 9$ &  old \\
\hline
0.09 & 0.0062/0.0186 & $28^3 \times 96$ & F6 & $950 \times 9$ & \texttt{new} \\
0.09 & 0.0124/0.031 & $28^3 \times 96$ & F4 & $1995 \times 9$ & \texttt{new} \\
0.09 & 0.0093/0.031 & $28^3 \times 96$ & F3 & $949 \times 9$  & \texttt{new} \\
0.09 & 0.0062/0.031 & $28^3 \times 96$ & F1 & $995 \times 9$  & old \\
0.09 & 0.00465/0.031 & $32^3 \times 96$ & F5 & $651 \times 9$ & \texttt{new} \\
0.09 & 0.0031/0.031 & $40^3 \times 96$ & F2 & $959 \times 9$  & \texttt{new} \\
0.09 & 0.0031/0.0186 & $40^3 \times 96$ & F7 & $701 \times 9$ & \texttt{new} \\
%0.09 & 0.0031/0.0031 & $40^3 \times 96$ & F8 & $576 \times 9$ & \texttt{new} \\
0.09 & 0.00155/0.031 & $64^3 \times 96$ & F9 & $790 \times 9$ & \texttt{new} \\
\hline
0.06 & 0.0072/0.018  & $48^3 \times 144$ & S3 & $593 \times 9$ & \texttt{new} \\
0.06 & 0.0054/0.018  & $48^3 \times 144$ & S4 & $582 \times 9$ & \texttt{new} \\
0.06 & 0.0036/0.018  & $48^3 \times 144$ & S1 & $749 \times 9$ & \texttt{update}\\
0.06 & 0.0025/0.018  & $56^3 \times 144$ & S2 & $799 \times 9$ & \texttt{new} \\
0.06 & 0.0018/0.018  & $64^3 \times 144$ & S5 & $572 \times 9$ & \texttt{new} \\
%0.06 & 0.0036/0.0108 & $64^3 \times 144$ & S6 & $600 \times 0$ & NA \\
\hline
0.045 & 0.0028/0.014 & $64^3 \times 192$ & U1 & $747 \times 1$ & old \\
%%%\hline\hline
\end{tabular}
\end{ruledtabular}
\end{table}
We use the MILC asqtad lattices listed in Table~\ref{tab:milc-lat}.
%
%Since the publication of Ref.~\cite{Bae:2011ff}, we have added
%eight new fine ensembles 
%(F2, F3, F4, F5, F6, F7, F8, and F9)
%and four new superfine ensembles
%(S2, S3, S4, and S5).
%
%In addition we have improved the statistics on ensemble S1.
%
As already noted, the eleven extra ensembles compared to
Ref.~\cite{Bae:2011ff} allow us to control the continuum and
sea-quark mass extrapolations with much greater confidence.
To give a sense of the range of these parameters, we present
in Table~\ref{tab:milc:para}
values for $a^2$, the light sea-quark pion mass ($\sqrt{L_P}$)
and the mass of the unphysical flavor-non-singlet $\bar s s$ state 
composed of sea quarks ($\sqrt{S_P}$). We will extrapolate/interpolate
to physical sea-quark masses using $L_P$ and $S_P$, while simultaneously
extrapolating $a^2$ to zero.
We exclude the coarse ensembles from Table~\ref{tab:milc:para} as
they are not used in the final extrapolation. 

\begin{table}[tb!]
\caption{Values for $a^2$, $\sqrt{L_P}$ and $\sqrt{S_P}$ on the
MILC ensembles used for our chiral-continuum extrapolation.
$a$ is determined from the mass-dependent values for
$r_1/a$ obtained by the MILC collaboration~\cite{Bazavov:2009bb,Bernard},
together with $r_1=0.3117\;$fm.
}
\label{tab:milc:para}
\begin{ruledtabular}
\begin{tabular}{ c | c | c | c }
ID & $a^2$(0.01fm${}^2$) & $\sqrt{L_P}$(MeV) & $\sqrt{S_P}$(MeV)\\
\hline
% SS: coarse lattices not used for chiral-continuum extrapolation
%ID & a m_l/a m_s&  a^2         &   L_P        & S_P
%
%C1 & 0.03/0.05  & 1.3835085796 & 0.4018609959 & 0.66886777 \\
%C2 & 0.02/0.05  & 1.3897948657 & 0.2714194105 & 0.66967218 \\
%C3 & 0.01/0.05  & 1.4175367279 & 0.1384060832 & 0.67142188 \\
%C4 & 0.007/0.05 & 1.3993049339 & 0.0993048401 & 0.67432189 \\
%C5 & 0.005/0.05 & 1.3866463792 & 0.0716261129 & 0.67782623 \\
\hline
F6 & 0.673 & 350 & 598 \\
F4 & 0.706 & 485 & 765 \\
F3 & 0.708 & 422 & 766 \\
F1 & 0.710 & 346 & 764 \\
F5 & 0.710 & 294 & 751 \\
F2 & 0.710 & 243 & 755 \\
F7 & 0.710 & 248 & 596 \\
%F8 & 0.693 & 257 & 257 \\
F9 & 0.710 & 174 & 759 \\
\hline
S3 & 0.348 & 440 & 694 \\
S4 & 0.348 & 383 & 694 \\
S1 & 0.346 & 314 & 696 \\
S2 & 0.347 & 262 & 695 \\
S5 & 0.348 & 222 & 691 \\
\hline
U1 & 0.192 & 316 & 703 \\
\end{tabular}
\end{ruledtabular}
\end{table}

We see from Table~\ref{tab:milc:para} that both fine and superfine
lattices have a substantial range of pion masses, % $\sqrt{L_P}$,
in the former case reaching down almost to the physical value.
The fine lattices also have two values for $\sqrt{S_P}$,
allowing interpolation to the ``physical'' value,
$0.6858\;$GeV~\cite{Davies:2009tsa}.
For the continuum extrapolation,
the relevant quantity for our action is $a^2$,
and we see that this varies by almost a factor of four.

Staggered fermions introduce an unwanted ``taste'' degree of freedom,
with each lattice staggered flavor giving rise to four 
degenerate tastes in the continuum limit.
This unwanted degeneracy is removed by the fourth-root prescription,
and we assume that this leads to the correct continuum limit.
The effect of this prescription, as well as that of using a mixed action,
can be incorporated into a chiral effective theory 
describing the staggered fermion formulation: staggered chiral
perturbation theory (SChPT)~\cite{Lee:1999zxa,
Aubin:2003mg,VandeWater:2005uq,Bae:2010ki}.
In particular, we use SU(2) SChPT at next-to-leading order (NLO)
to obtain the functional
form needed for extrapolations in light quark masses and $a^2$.

\section{Data analysis and fitting}
\label{sec:fit}

We calculate $B_K(1/a,\text{NDR})$ on each ensemble
following the method explained in Ref.~\cite{Bae:2010ki}, using
multiple measurements on each configuration.
For the valence $d$ quarks we use the four masses (in lattice units)
$am_x=am_s^{\rm nom}\times \{0.1,0.2,0.3,0.4\}$,
where $m_s^{\rm nom}$ is a nominal strange quark mass which lies
fairly close to the physical value. 
For coarse, fine, superfine and ultrafine ensembles we take
$am_s^{\rm nom}=0.05$, $0.03$, $0.018$ and $0.014$, respectively.
For the valence $s$ quarks we use the three masses
$am_y=am_s^{\rm  nom}\times \{0.8,0.9,1.0\}$.
With these values we are extrapolating to the
``physical'' $\bar d d$ mass
($158\;$MeV---the mass of a flavor-non-singlet $\bar d d$ state)
using lattice valence pions with masses in the range $\sim 200-400\;$MeV.
For the valence $s$ quark we are extrapolating from the range
$\sim 550-620\;$MeV to the ``physical'' value $686\;$MeV.
Thus both valence extrapolations are relatively short.
In addition, since $m_x/m_y \le 1/2$ and 
$m_y\sim m_s^{\rm phys}$, they are done in a regime where SU(2) ChPT should
be valid.

We perform the chiral and continuum extrapolations in three stages.
First, the valence $d$ quark is extrapolated to $m_d^{\rm phys}$, using a 
functional form based on NLO SChPT, although including higher-order
analytic terms. Details are as in Ref.~\cite{Bae:2011ff}.
This ``X-fit'' is done separately on each ensemble, 
using a correlated fit with Bayesian priors.
We correct for non-analytic contributions from  
taste breaking in valence and sea pions 
and from the unphysical value of $m_\ell$
using the one-loop chiral logarithms (which are predicted
without unknown constants).
This correction depends on the ensemble, and ranges in size
from $3-5\%$.
The X-fits on all the new ensembles are very similar to those
displayed in Refs.~\cite{Bae:2010ki,Bae:2011ff,Bae:2013lja}.

The second stage, or ``Y-fit'', is the extrapolation from our three values of
$m_y$ to $m_s^{\rm phys}$.
The dependence on $m_y$ is expected to be analytic, and we find that
a linear fit works well. Examples of such fits are shown in
Ref.~\cite{Bae:2010ki}; those on the new ensembles are very similar.

The third and final extrapolation is that in $m_\ell$, $m_s$
and $a^2$. Here we improve on Refs.~\cite{Bae:2010ki,Bae:2011ff}
both by having a larger range of lattice parameters and
by doing these extrapolations simultaneously.
As in these earlier works, we find that we cannot
obtain good fits if we include results from the coarse lattices
and so exclude them. 

We now describe the fit functions and fitting approach
used in the simultaneous chiral-continuum extrapolation.
Our first fit function assumes that the lattice operator is 
perfectly matched to that in the continuum.
Then SU(2) SChPT at NLO predicts a linear
residual dependence on $m_\ell$ and $a^2$ 
(after taste-breaking in the chiral logarithms has been removed 
by hand).
The only constraint on the $m_s$ dependence is that it must be analytic,
but for our short interpolation or extrapolation it is likely to be
well described by a linear dependence.
The appropriate fit function is thus
\begin{equation}
\label{eq:fit1}
f_1(a^2,L_P,S_P) = c_1 + c_2 (a\Lambda_Q)^2 + c_3 \frac{L_P}{\Lambda_X^2} 
+ c_4 \frac{S_P}{\Lambda_X^2} \,,
\end{equation}
where we are using $L_P$ and $S_P$ as stand-ins for $m_\ell$ and
$m_s$, respectively.
Taking $\Lambda_Q = 0.3\;$GeV and $\Lambda_X = 1.0$ GeV
(i.e. respectively a typical QCD scale and chiral expansion scale)
we expect $c_2-c_4= {\cal O}(1)$.
We stress that Eq.~(\ref{eq:fit1}) is only valid for a small
range of $S_P$ around the physical value. In particular,
the linear dependence on $S_P$ is not assumed or expected to remain
valid down to $S_P=0$, and there is no {\em a priori} expectation that 
$c_4\approx c_3$. The latter relation would only hold were we in
the regime where SU(3) ChPT at NLO was valid.

\begin{table}[tb!]
\caption{Fit functions and quality.}
\label{tab:fit-type}
\begin{ruledtabular}
\begin{tabular}{ c | l | l | c}
%%%\hline\hline
fit type & fit function & Constraints & $\chi^2/\text{d.o.f}$ \\
\hline
B1 & $1 + a^2 + L_P + S_P$ & 2, 3, 4 & 1.48 \\ 
B2 & $\text{B1} + a^2 L_P + a^2 S_P$ & 2, 3, 4, 5, 6 & 1.47\\
B3 & $\text{B1} + \alpha_s^2 + a^2 \alpha_s + a^4$ & 2, 3, 4, 7, 8, 9 & 1.47\\
B4 & $\text{B2} + \alpha_s^2 + a^2 \alpha_s + a^4$ & 2, 3,$\ldots$, 9 & 1.47\\
\hline
N1 & same as B1 & none & 1.91 \\
%%%
% SS: dropped N2 as too complicated to explain rationale and
%     adds little to discussion
%%%
%
%N2 & same as B2 & none & 2.04 \\
%%%\hline\hline
\end{tabular}
\end{ruledtabular}
\end{table}

When using the form $f_1$ we either 
apply Bayesian constraints, $c_i  = 0 \pm 2$, to $c_2$, $c_3$ and $c_4$
(fit B1)
or leave all four coefficients free (fit N1).
We consider the range $\pm 2$ to be a fairly conservative
choice for the constraints.
These details, along with the resulting quality of fit,
are collected in Table~\ref{tab:fit-type}.
The parameters of both fits turn out to be almost identical
and are given in Table~\ref{tab:fit-param}.
Both fits are reasonable, with the sizes of the 
constrained coefficients $c_2-c_4$
well within the expected range of $\pm 2$.
Thus the constraints are not important for this fit.\footnote{%
Despite having almost identical fit parameters,
the $\chi^2/\text{d.o.f.}$ values for the two fits differ
because in a Bayesian fit one augments both $\chi^2$
and the effective number of data points~\cite{Lepage:2001ym}.}

\begin{table}[tb!]
\caption{Parameters of representative fits.}
\label{tab:fit-param}
\begin{ruledtabular}
\begin{tabular}{ c | c  c c  c c }
%%%\hline\hline
fit type & $c_1$  & $c_2$  & $c_3$ & $c_4$ &  \\
\hline
%%% SS: converted c2, c5, c6, c8 by
%%%     multiplying by (0.197 GeV fm/(0.3GeV*0.1fm))^2=43.12
%%%     and converted c9 by multiplying by (43.12)^2
%%%     c1 and c7 are unchanged
B1/N1 & 0.542(7) & 0.7(3) & -0.17(1) & 0.00(1) & \\
B4 & 0.54(1) & 0.3(4)  & -0.17(2) & -0.00(2) & \\
\hline
fit type & $c_5$  & $c_6$  & $c_7$ & $c_8$ & $c_9$ \\
\hline
B4 & -0.2(8) & 0.2(1) & 0.1(2) & 0.1(2) & 0.01(3) \\
%%%\hline\hline
\end{tabular}
\end{ruledtabular}
\end{table}

\begin{figure}[tb!]
\centering
\includegraphics[width=20pc]{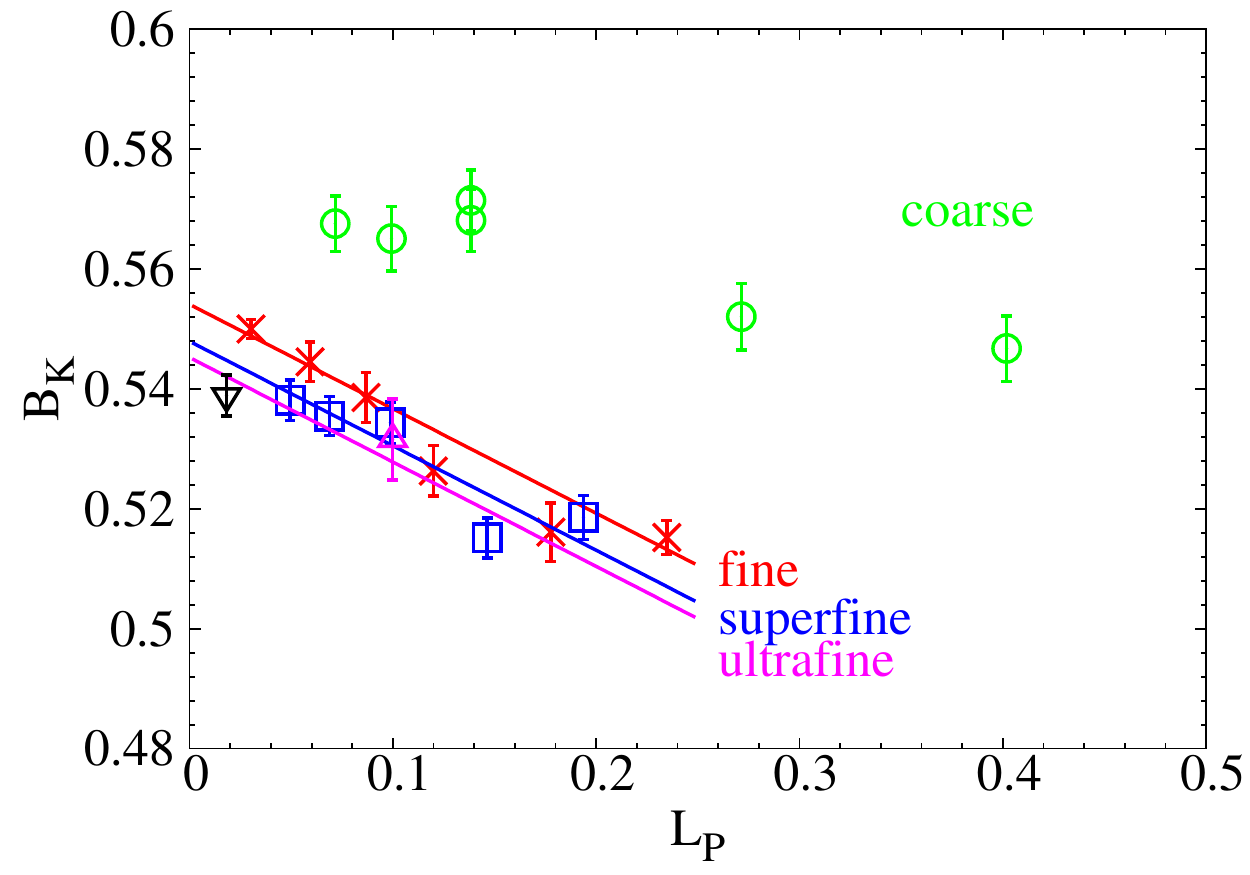}
\caption{$B_K(2\,\text{GeV},\text{NDR})$ vs. $L_P \, (\text{GeV}^2)$,
  with the B1 fit. The N1 fit is indistinguishable. The result of the
  extrapolation is shown by the (black) triangle.  The fit function
  for the superfine ensembles is plotted using the average values of
  $a^2$ and $S_P$, while for the fine ensembles the average of the
  values on ensembles $F1$, $F2$ and $F4$ is used.  Coarse lattice
  results are shown for comparison; they are not included in the fit.
  Results from ensembles F6 and F7 are not shown (as explained in the
  text), but are included in the fit.
}
\label{fig:bk:B1}
\end{figure}
The quality of the resulting fit is illustrated by Fig.~\ref{fig:bk:B1}.
A complication in displaying the fit is that within the fine ensembles
there is a range of values of $a^2$ and $S_P$ 
(and similarly for the superfine ensembles)
and this feature cannot be displayed in a two-dimensional plot.
The fit lines shown are for average values of $a^2$ and $S_P$, and,
even with a perfect fit, would not pass exactly through the corresponding
points.
As can be seen from Table~\ref{tab:milc:para}, this is a small effect
except for ensembles F6 and F7, which have significantly smaller values of
$S_P$. Thus these two ensembles are not included in the figure
(although they are included in the fit itself).
The results on the coarse ensembles are also shown, although they are
not included in the fit. It is clear that the slope versus
$L_P$ is significantly different on the coarse ensembles, and also that there
are large discretization errors. These are the features that make fits
including the coarse ensembles unstable, as high order terms are needed
to include them.

\begin{figure}[tb!]
\centering
\includegraphics[width=20pc]{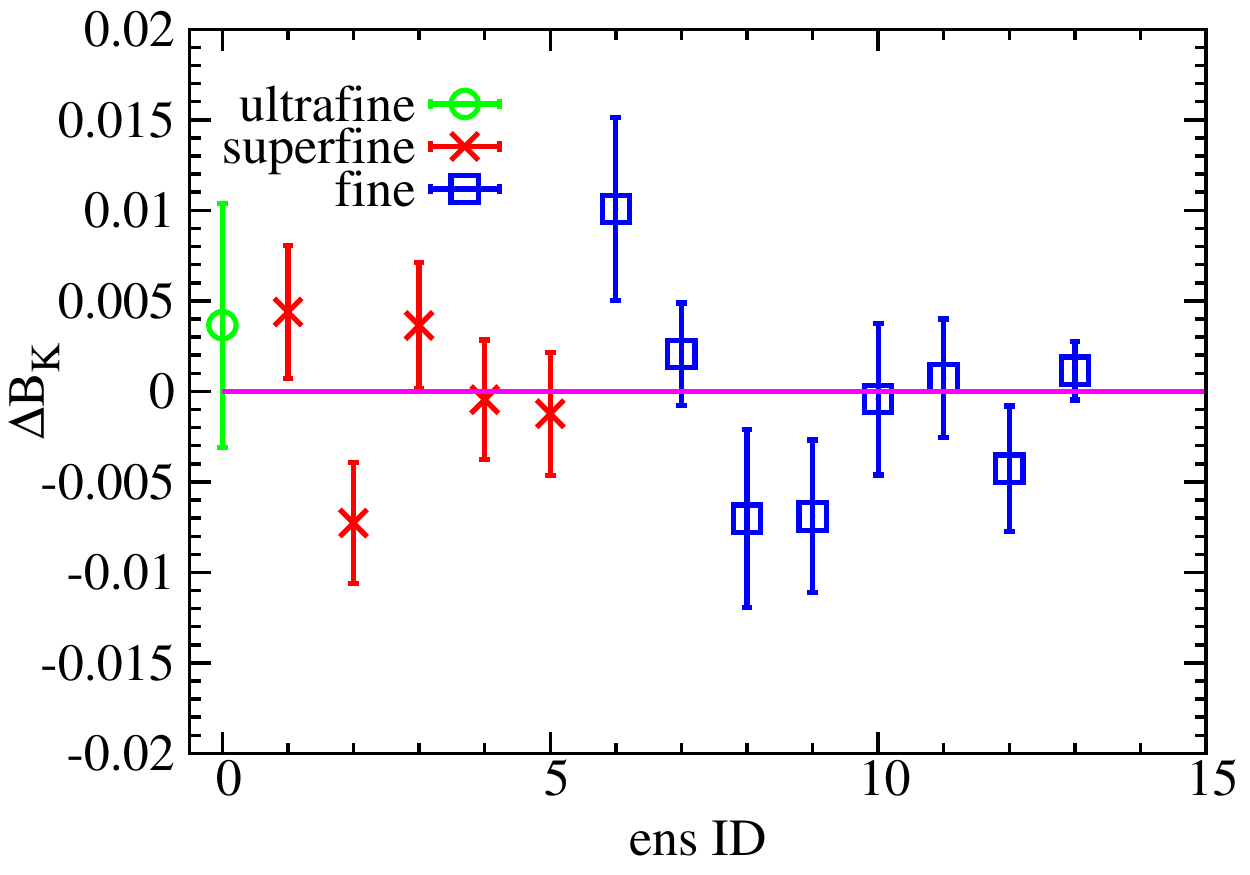}
\caption{Residuals $\Delta B_K$ for fit B1.
The superfine lattices are ordered by decreasing
$L_P$ (S3, S4, S1, S2 and S5), while the fine lattices
are ordered F6, F4, F3, F1, F5, F2, F7 and F9.
These are the orders used in Tables~\ref{tab:milc-lat} 
and \ref{tab:milc:para}.
}
\label{fig:Deltabk:B1}
\end{figure}

The complication of having different values of
 $a^2$ and $S_P$ can be avoided by considering
the residuals $\Delta B_K(i)=B_K(i)-f_1(i)$, where $i$ labels the ensembles.
These are shown in Fig.~\ref{fig:Deltabk:B1}.

As already noted,
the values of the parameters $c_2-c_4$ lie in the expected
range. In particular, if one writes the $c_1$ and $c_2$ terms
in the form
$c_1 [1 + (a\Lambda)^2 ]$ then
we find $\Lambda\approx 350\;$MeV, which is a reasonable
scale for a discretization error.
We also note that the SU(3) symmetry relation $c_3=c_4$ does
not hold. Indeed, we find no significant dependence on $S_P$
in the vicinity of the strange quark mass.

We now turn to our other fits. SU(2) SChPT
is a joint expansion in $a^2$ and $L_P$ (ignoring possible
factors of $\alpha$ multiplying $a^2$), so at NNLO we expect
a term of the form $c_5 (a\Lambda_Q)^2 (L_P/\Lambda_X^2)$ 
with coefficient $c_5\sim {\cal O}(1)$.
In fit B2 we include this term, as well as its SU(3) counterpart
$c_6 (a\Lambda_Q)^2 (S_P/\Lambda_X^2)$, 
with coefficients constrained as for $c_2-c_4$.
In this case we find that the constraints are needed to obtain sensible fits.
The resulting fit, however, lies very close to B1, and, as shown in
Table~\ref{tab:fit-type}, does not have an improved $\chi^2/\text{d.o.f.}$.
This simply reflects the fact that our data has similar slopes versus $L_P$
on the fine and superfine lattices.

We next consider the impact of operator matching errors on the fit function.
Since we use one-loop matching,\footnote{%
In fact, we do not do a complete one-loop matching, since we keep only
those lattice four-fermion operators composed of bilinears
having the same taste as the external pions~\cite{Bae:2010ki},  
Thus there {\em are} matching corrections proportional to $\alpha$. 
These, however, appear only at NNLO in SU(2) SChPT~\cite{Bae:2010ki}, 
and are of the form $\alpha a^2$ or $\alpha L_P$.
Furthermore, the numerical coefficients of these terms are 
small~\cite{KLS2}. Thus we choose to treat them as effectively
of NNNLO, and do not include them in the fits.}
these errors are proportional to $\alpha^2$
(with $\alpha$ evaluated as the scale $1/a$).
Thus we also consider fits with a $c_7\alpha^2$ term, as well
as terms $c_8(a\Lambda_Q)^2\alpha$ and 
$c_9(a\Lambda_Q)^4$ arising from higher-order discretization errors. 
When adding these terms we find that Bayesian constraints
(for which we use $c_7-c_9=0\pm 2$)
are needed for stable fits. 
We have considered both fits in which only these three terms
are added to $f_1$ (fit B3) and in which all the terms described
above are included (fit B4).
The full list of fits we use is shown in Table~\ref{tab:fit-type}.

We find again that adding higher-order terms does not improve the
quality of the fits (see Table~\ref{tab:fit-type}), and also has
little impact on the resulting fit parameters.  For example, as shown
in Table~\ref{tab:fit-param}, the terms common between fits B1/N1 and
B4 are almost identical, while the extra terms in B4 are all small and
consistent with zero.
%-----------------------------
% The lack of change between the two fits is also
% seen by comparing the residuals for B4, shown in
% Fig.~\ref{fig:Deltabk:B4}, with those for B1 in
% Fig.~\ref{fig:Deltabk:B1}.
%------------------------------

In summary,  our data can be described well by the simple form $f_1$,
but is also consistent with the extra terms as long
as they have small coefficients.
In light of this, we have not extended the fits to
include the other possible NNLO terms, e.g. that proportional
to $L_P^2$.
We also conclude that it is reasonable to use fit B1 for our central
value, while quoting the maximum difference between the results from
B1 and \{B2, B3, B4\} fits (which turns out to be
for fit B4) as an extrapolation systematic.

%
%\begin{figure}[tb!]
%\centering
%\includegraphics[width=20pc]{dbk_su2_g4_a2g2_a4_lp_sp_a2lp_a2sp_3pt}
%
%\caption{Residuals $\Delta B_K$ for fit B4.
%Notation as in Fig.~\ref{fig:Deltabk:B1}.
%{\bf SS: Is there any point in showing this figure? It looks very similar
%to the previous one.}
%
%}
%\label{fig:Deltabk:B4}
%\end{figure}
%

\section{Error Budget} 
\label{sec:err-budget}
We present the error budget in Table~\ref{tab:err-budget}.
The largest error is from using one-loop matching,
which we estimate to be
$\Delta B_K/B_K=\alpha^2$, with $\alpha$ evaluated at scale $1/a$
for the ultrafine lattice.
This error is unchanged from Ref.~\cite{Bae:2011ff}.
In principle, one can determine the size of the $\alpha^2$
contribution from the chiral-continuum fit, given a sufficiently
extensive set of lattice ensembles. Indeed, this term is included in
fit B4 (with coefficient $c_7$). However, it is clear from the results
of the previous section that we do not have enough ensembles to pin
down $c_7$, especially given the large number of parameters in the
fit.  Thus, although fit B4 finds a small value, $c_7=0.1(2)$, we do
not think this is sufficiently reliable to take at face value, and
prefer the conservative approach of taking $c_7=\pm 1$ to estimate the
matching error.

The errors from continuum and sea-quark mass extrapolations have
been significantly reduced compared to Ref.~\cite{Bae:2011ff}. 
Previously, these errors were separate, and were estimated to be
1.9\% from the continuum extrapolation, 
1.5\% from the $am_\ell$ extrapolation, and
1.3\% from the $a m_s$  extrapolation~\cite{Bae:2011ff}. 
The combined error was thus 2.8\%.
The addition of the new ensembles and the use of a combined extrapolation
has reduced this error to 0.9\%.

The statistical error is essentially unchanged
from Ref.~\cite{Bae:2011ff}.
As in that work, we have not accounted for the impact of auto-correlations.
We do observe about a 20\% increase in the statistical error due to 
auto-correlations, as reported in Ref.~\cite{Yoon:2009th}.
However, since there is significant uncertainty in the size of
this increase, and 
since this effect is much smaller than our current systematic errors,
we have decided to neglect it in this paper.
%
%However, were the total systematic error to be reduced down to the
%sub-percent level, then we would need to take into account this
%autocorrelation effect in our statistical error.
%

The error from the X-fits is estimated similarly to
the approach used in Ref.~\cite{Bae:2011ff},
namely by considering the effects of doubling the 
Bayesian priors and of switching to the eigenvalue-shift method of 
fitting~\cite{Jang:2011fp}.
We find that neither of these changes lead to statistically
significant shifts in $B_K$ on any ensemble. If we combine the
two fractional shifts in quadrature we find an $\approx 0.1\pm 0.1\%$
shift on essentially all fine and superfine ensembles.\footnote{%
The combined shift is $0.4\pm 0.5\%$ on the U1 ensemble.
The larger error 
is due to our smaller number of measurements on this ensemble.}
We thus take this as our estimate of this (very small) effect.

The error in Y-fits arises from the uncertainty in the
functional form used to extrapolate in the valence strange-quark mass.
$m_y$. We have used linear fits for our central value, but cannot
rule out a small quadratic component. Thus we have repeated the entire
analysis using quadratic Y-fits, finding a statistically significant
2.0\% downward shift in the final value of $B_K$. This we quote as
the corresponding systematic error.
This error is much larger than that quoted in Ref.~\cite{Bae:2011ff},
but we think the present estimate is both more conservative and 
more reliable.

The finite-volume error is estimated as in 
Refs.~\cite{Bae:2011ff,Kim:2011qg}.
For our central value we do the X-fits with SChPT expressions including
the (analytically known) finite-volume corrections to the chiral logarithms.
We then repeat the entire analysis using infinite-volume chiral logarithms 
in X-fits, and take the difference between the resulting values
of $B_K$ as an error estimate. The rationale for this choice
is that one-loop chiral 
logarithms typically provide only a semi-quantitative estimate of the
size of finite-volume effects.
The resulting error is numerically small.

The final two errors are those due to uncertainty in the choices of scale and
of the appropriate value of the pion decay constant to use in the 
chiral logarithms entering X-fits. 
We follow the MILC collaboration and set the scale using 
$r_1=0.3117(22)\;$fm~\cite{Bazavov:2011aa}.
To obtain our central value of $B_K$ we take $r_1=0.3117$.
We then repeat the analysis on each ensemble 
using both $r_1=0.3139$ and $0.3095$. 
We find that both changes lead, on all fine, superfine and
ultrafine ensembles, to shifts of magnitude $\approx 0.3\%$ in $B_K$.
Given this uniformity, we expect a similar shift in the final
answer and thus take this as our error estimate.

For the decay constant we use $f_\pi=132\;$MeV 
(a somewhat outdated approximation to the physical value of $130.41\;$MeV)
for determining the central value of $B_K$,
and then repeat the analysis using the decay constant in the
SU(2) chiral limit, $f_\pi^{(0)}=124.2\;$MeV~\cite{Bazavov:2009bb}.
This is the same procedure as in Refs.~\cite{Bae:2010ki,Bae:2011ff}.
In this case the shift in $B_K$ does vary significantly between ensembles,
so we repeat the entire analysis using both values of $f_\pi$, 
and take the difference in final values as our estimate of the error.
As the Table shows, the resulting error is very small.

\begin{table}[tbp]
\caption{Error budget for $B_K$ using SU(2) SChPT fitting.
  \label{tab:err-budget}}
\begin{ruledtabular}
\begin{tabular}{ l | l l }
  cause & error(\%) & memo \\
  \hline
  STATISTICS         & 0.64  & jackknife \\
\hline
  matching factor    & 4.4   & see text \\
  $ \left\{ \begin{array}{l} 
    \text{discretization} \\ a m_\ell \text{ extrap} \\ a m_s \text{ extrap}
  \end{array} \right\} $
                     & 0.9  & diff.~of B1 and B4 fits \\
  X-fits             & 0.1   & see text \\
  Y-fits             & 2.0   & diff.~of linear and quad. \\
  finite volume      & 0.4  & diff.~of $V\!=\!\infty$ and FV fit \\
  $r_1$              & 0.3  & $r_1$ error propagation \\
  $f_\pi$            & 0.1  & $132\;$MeV vs. $124.4\;$MeV\\
\hline
  TOTAL SYSTEMATIC   & 4.9 &
\end{tabular}
\end{ruledtabular}
\end{table}
\section{Conclusion}
\label{sec:conclude}

Our final results are
\begin{align}
& B_K(2\;{\rm GeV},\text{NDR}) & = & \ 0.5388 \pm 0.0034 \pm 0.0266  
\\
& \hat{B}_K & = & \ 0.7379 \pm 0.0047 \pm 0.0365 
\label{eq:BKhatfinal}
\end{align}
where the first errors are statistical and the second systematic.
$\hat{B}_K = B_K (\text{RGI})$ is the renormalization group
invariant value of $B_K$.
This result  supersedes our previous result,
$\hat{B}_K = 0.727 \pm 0.004 \pm 0.038$~\cite{Bae:2011ff},
with which it is completely consistent.
Although the changes are numerically small, they are significant.
By adding many new ensembles we now can properly extrapolate
in sea quark masses (rather than estimate the effect of such
an extrapolation and include it as an error). This is the main reason
for the small increase in the central value.
It also leads to a significant reduction in the systematic errors
from the continuum and sea-quark mass extrapolations.
On the other hand, a more careful estimate of the systematic
error in the valence strange-quark mass extrapolation has led
to a significant increase in this error.
All told, the overall error is only slightly reduced, but, more
importantly, the methods of estimating errors have been improved.

Our result (\ref{eq:BKhatfinal}) is consistent with the world average
presented in Ref.~\cite{Aoki:2013ldr}, $\hat{B}_K=0.766(10)$.  Our
error is larger than this average primarily because of our use of
one-loop matching factors.
We are presently working on obtaining the matching factors using
non-perturbative renormalization~\cite{Martinelli:1994ty}, which
should result in substantial reduction of the matching
error\cite{Kim:2012ng,Lytle:2013qoa}.
In addition, we plan to calculate the matching factors perturbatively
at the two-loop level using automated perturbation theory.

Our second-largest error is that from Y-fits. This error can, however,
be essentially removed in a straightforward way by using valence
strange quarks tuned to the physical value.

\begin{comment}
The result for $\hat{B}_K$, when combined with recent updates on
exclusive $V_{cb}$, gives a theoretical estimate for $\varepsilon_K$
from the Standard Model which differs by $3-4 \sigma$ from experiment.
%
Details of this analysis will be reported in a separate paper.
% 
\end{comment}

% EDIT

\begin{acknowledgments}
We are grateful to Claude Bernard and the MILC collaboration
for private communications.
C.~Jung is supported by the US DOE under contract DE-AC02-98CH10886.
The research of W.~Lee is supported by the Creative Research
Initiatives program (2013-003454) of the NRF grant funded by the
Korean government (MSIP).
W.~Lee would like to acknowledge the support from KISTI supercomputing
center through the strategic support program for the supercomputing
application research [No. KSC-2012-G2-01].
The work of S.~Sharpe is supported in part by the US DOE grant
no.~DE-FG02-96ER40956.
Computations for this work were carried out in part on QCDOC computers
of the USQCD Collaboration at Brookhaven National Laboratory and in
part on the DAVID GPU clusters at Seoul National University.
The USQCD Collaboration are funded by the Office of Science of the
U.S. Department of Energy.
\end{acknowledgments}

%-----------
% reference
%-----------
% SS: commented out to get the proper hyperlinks in references
%\bibliographystyle{apsrev} %%% physical review
\bibliography{ref} %%% ref.bib file
\end{document}